\documentclass[conference]{IEEEtran}
\IEEEoverridecommandlockouts

\usepackage{cite}
\usepackage{amsmath,amssymb,amsfonts}
\usepackage{algorithmic}
\usepackage{graphicx}
\usepackage{textcomp}
\usepackage{xcolor}
\usepackage{multirow}
\usepackage{hyperref}
\usepackage{array}

\newcolumntype{C}[1]{>{\centering\arraybackslash}m{#1}}
\newcolumntype{L}[1]{>{\raggedright\arraybackslash}m{#1}}
\newcolumntype{J}[1]{>{\arraybackslash}m{#1}}
\def\BibTeX{{\rm B\kern-.05em{\sc i\kern-.025em b}\kern-.08em
    T\kern-.1667em\lower.7ex\hbox{E}\kern-.125emX}}
\begin{document}

\title{Evaluating Social Engineering Risks in AI-based Interaction using Biometrics and a Gaming Setup
\thanks{Support by Cátedra ENIA UAM-VERIDAS en IA Responsable (NextGenerationEU PRTR TSI-100927-2023-2), M2RAI (PID2024-160053OB-I00, MICIU/FEDER), TRUST-ID (PID2025-173396OB-I00, MICIU/AEI/EU) and PowerAI+ (SI4/PJI/2024-00062, CM/UAM). J.I. is an FPI MINECO/FEDER.}}
\author{
\IEEEauthorblockN{
Roberto Daza\IEEEauthorrefmark{1}\IEEEauthorrefmark{2},
Javier Irigoyen\IEEEauthorrefmark{1},
Ivan Lopez\IEEEauthorrefmark{1},
Raquel Rodriguez-Carvajal\IEEEauthorrefmark{3},\\
Laura Gomez-Carbajo\IEEEauthorrefmark{3},
Julian Fierrez\IEEEauthorrefmark{1},
Ruben Tolosana\IEEEauthorrefmark{1}, and
Aythami Morales\IEEEauthorrefmark{1}\IEEEauthorrefmark{4}
}

\IEEEauthorblockA{
\IEEEauthorrefmark{1}\textit{BiometricsAI, Universidad Autónoma de Madrid (UAM), Spain}
}

\IEEEauthorblockA{
\IEEEauthorrefmark{2}\textit{GHIA, Universidad Autónoma de Madrid (UAM), Spain}
}

\IEEEauthorblockA{
\IEEEauthorrefmark{3}\textit{Biological and Health Psychology Department, Universidad Autónoma de Madrid (UAM), Spain}
}

\IEEEauthorblockA{
\IEEEauthorrefmark{4}\textit{Universidad de Las Palmas de Gran Canaria (ULPGC), Spain}
}

\IEEEauthorblockA{
Corresponding author: roberto.daza@uam.es
}
}



\maketitle

\begin{abstract}

We introduce AIriskEval-gaming, an open platform and dataset to evaluate social engineering risks in LLM-mediated multimodal interaction through controlled games. It supports human--human, human--AI and AI--AI settings, combining configurable game templates, role-conditioned LLM agents, psychology-informed participant profiling, structured interaction trees, and synchronized behavioral and biometric acquisition, filtering, and deep-learning-based feature extraction. The dataset (AIriskEval-gaming-db) was collected from 15 participants who interacted with a role-conditioned GPT-5.4 agent in two concatenated games: an adapted Prisoner's Dilemma and an Ultimatum Game. It comprises 340 GB of raw and processed multimodal data across six streams: interaction logs, video, screen recordings, gaze logs, smartwatch signals, and game/questionnaire metadata. These data include interaction paths, written justifications, psychological profiles, subjective feedback, perceived counterpart identity, game outcomes, and derived behavioral, facial, and gaze features. Alongside the dataset, we provide descriptive analyses characterizing the resulting multimodal data. Rigorous risk evaluation is essential for the deployment of secure AI systems, as it enables the identification and mitigation of vulnerabilities, ensures the protection of sensitive data, and supports compliance with evolving regulatory and ethical standards in society. The dataset and related code are available on GitHub.
\end{abstract}

\maketitle

\section{Introduction}

In the cybersecurity landscape, the human factor remains a vulnerability, particularly in social engineering scenarios. Traditionally, social engineering attacks have relied on static deception strategies, such as phishing templates, impersonation, or hand-crafted pretexts. However, the emergence of Generative Artificial Intelligence (GenAI) and Large Language Models (LLMs) is shifting this threat model towards more adaptive and personalized forms of influence~\cite{bi2026feasibility}.

LLM-based agents are no longer limited to passive text generation. They can participate in social exchanges, support decision-making, negotiate, persuade, and adapt their communication style to the user and the interaction context. Recent studies show that their behavior can be modulated by characterization, previous interaction history, personalization, and role-based attribution, making it possible to instantiate agents with different apparent preferences, identities, or strategies~\cite{mei2024turing,dvorak2025adverse,bi2026feasibility}. This creates a new human-centered risk: the vulnerable component is not only the software, but also the human decision process. This risk can be studied using social decision-making games, which provide a controlled framework capable of quantifying abstract social constructs. Recent behavioral research leverages these games from two perspectives: 1) how LLMs behave in tasks that involve trust, fairness, and cooperation~\cite{mei2024turing}; and 2) how humans respond to counterparts whose decisions are human-made, AI-mediated, or not clearly disclosed as human or AI~\cite{dvorak2025adverse}. 

However, decisions, interaction logs, and questionnaires are insufficient to fully characterize human responses during social decision-making. Behavioral and biometric signals can provide complementary real-time evidence of cognitive and  affective states, which is valuable to audit human susceptibility to AI-mediated influence. Although multimodal signals (e.g. facial biometrics, gaze, EEG, heart-rate, and interaction dynamics) have been used in human--computer interaction~\cite{daza2023edbb,daza2024improveimpactmobilephones,daza2025smartevr}, to detect stress in cybersecurity scenarios~\cite{almehmadi2025cyber}, and to analyze trust towards artificial agents~\cite{cominelli2021promises}, their integrated application for auditing AI-driven social engineering remains scarce.


\begin{figure*}
  \centering
  \includegraphics[width=0.98\textwidth]{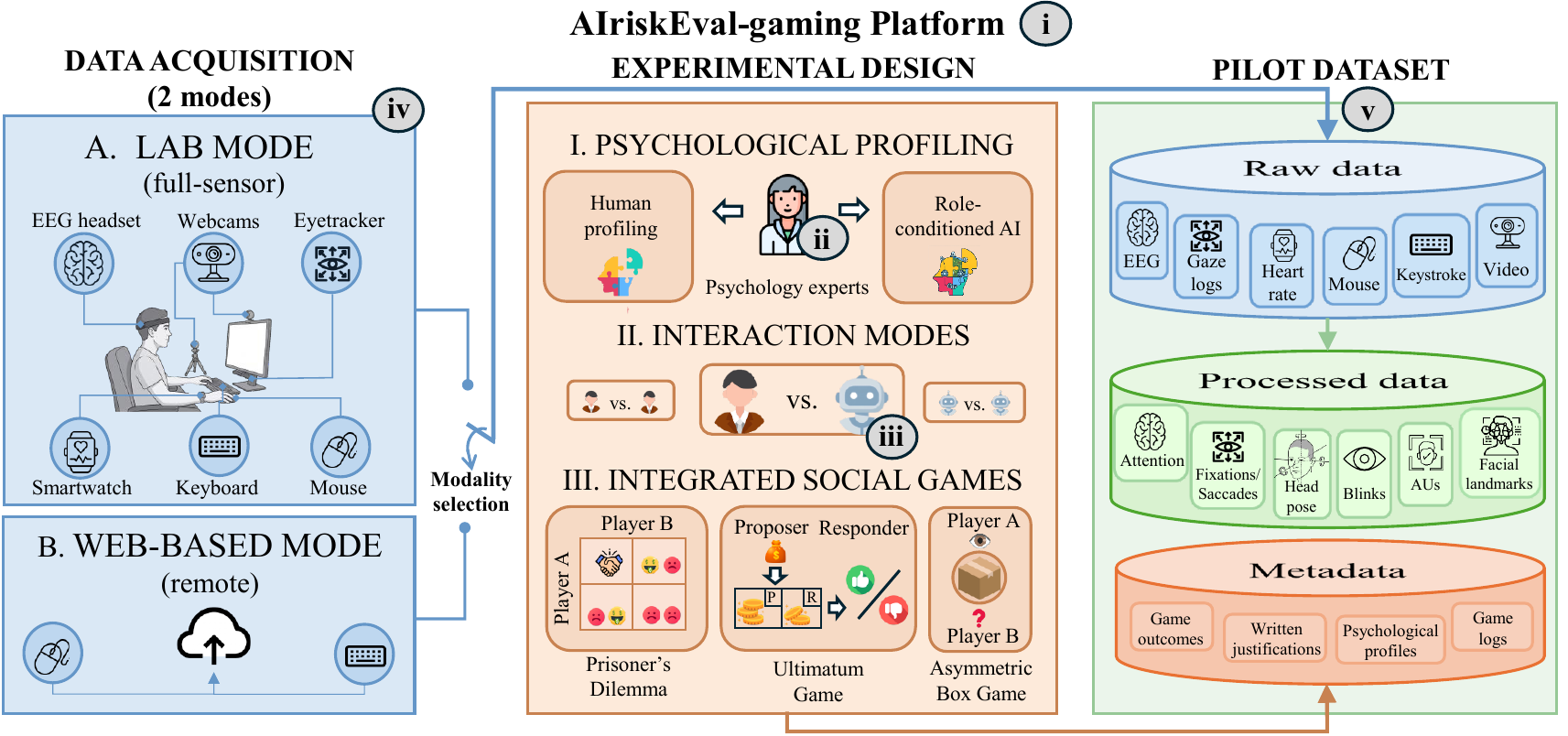}
  \caption{Overview of the AIriskEval-gaming platform and dataset, illustrating the main contributions of the paper: (i) the AIriskEval-gaming platform; (ii) psychology-informed profiling for human profiles and role-conditioned AI agents; (iii) interaction modes and social decision-making games; (iv) web-based and full-sensor acquisition; and (v) the dataset, organized into raw data, processed data, and metadata for auditing adaptive social engineering risks.}
  \label{fig:Overview}
\end{figure*}

In this scenario, we have developed AIriskEval-gaming,\footnote{ \url{https://github.com/BiometricsAI/AIriskEval-gaming}} a new platform and research line aimed at overcoming the current limitations of risk identification and evaluation in human--AI social decision-making:

\begin{itemize}
    \item To our knowledge, few platforms are designed to study these games from a security perspective, lacking comparable human--human, human--AI, and AI--AI settings to isolate AI-driven social engineering risks. 

    \item The influence of psychologically grounded AI roles on human trust and cooperation remains underexplored, especially with respect to how user personality traits shape interactions with these agents.

    \item Existing datasets focus mainly on text logs and final game choices~\cite{mei2024turing,akata2025playing,dvorak2025adverse}, lacking the synchronized behavioral and biometric signals required to audit cognitive and affective states in real-time. 
\end{itemize}

These limitations hinder the development of reproducible resources for auditing adaptive social engineering risks in LLM-mediated interaction.

To address them, we introduce AIriskEval-gaming, a platform and dataset for studying role-based social engineering risks in social games involving LLM agents. The main contributions are shown in Fig.~\ref{fig:Overview}:
i) We present AIriskEval-gaming, a security-oriented experimental platform supporting human--human, human--AI, and AI--AI settings in both web-based and full-sensor laboratory modes.
ii) We design, with psychology experts, a protocol that uses shared psychological constructs to define LLM agent roles and profile participants, complemented by pre- and post-interaction questionnaires to assess trust, fairness, perceived identity, cooperation, and role effects.
iii) We define role-conditioned LLM agents and structured interaction trees that specify participant response options and adaptive AI replies across sequential game rounds.
iv) We integrate a synchronized multimodal pipeline aligning contextual data  (logs, conversations, questionnaires) with behavioral biometrics (face, gaze, keystroke and mouse dynamics) and physiological signals (EEG, heart rate, stress and inertial). 
v) We publish a multimodal dataset (AIriskEval-gaming-db, see the footnote on the first page) to study human behavior and interaction patterns with role-conditioned LLM agents from a risk analysis perspective considering social engineering vulnerabilities.


\section{Related Work}
\label{s:related-work}

Recent research highlights a shift from static malicious content towards adaptive AI-mediated influence. GenAI has been studied in digital deception and phishing~\cite{schmitt2024digital},  AI-generated spear-phishing can reach human-expert performance~\cite{heiding2026evaluating}, and frameworks such as SEAR demonstrate role-based context-aware social engineering~\cite{bi2026feasibility}. In parallel, social decision-making games have been used to evaluate LLM behavior~\cite{mei2024turing, akata2025playing}, human responses to AI-mediated decisions~\cite{dvorak2025adverse}, and role-conditioned agents induced by prompting~\cite{phelps2025machine}. However, these works focus on final decisions, model output, or game histories rather than real-time human states.

Existing experimental platforms, such as oTree, nodeGame, Empirica, and LIONESS, support behavioral experiments, multi-stage interactions, programmable bots, and economic or social decision-making games~\cite{chen2016otree,balietti2017nodegame,almaatouq2021empirica,giamattei2020lioness}. However, they do not integrate role-conditioned LLM agents or synchronized biometric acquisition. Similarly, while multimodal datasets from robotics, affective computing, and cybersecurity show the value of physiological signals for modeling user states~\cite{newman2022harmonic,heinisch2024physiological,bussolan2025multiphysio,almehmadi2025cyber}, they are not tailored to human--AI social decision-making games. To our knowledge, AIriskEval-gaming is the first security-oriented framework to integrate role-conditioned LLM agents, human psychological profiling, and synchronized multimodal signals (including biometric signals~\cite{daza2023edbb,daza2024improveimpactmobilephones,daza2025smartevr}), to evaluate social engineering risks.



\section{AIriskEval-gaming Platform Design}

\subsection{Overview}


AIriskEval-gaming is designed to support controlled social decision-making experiments through four components:

\begin{figure*}[t]
  \centering
  \includegraphics[width=\textwidth]{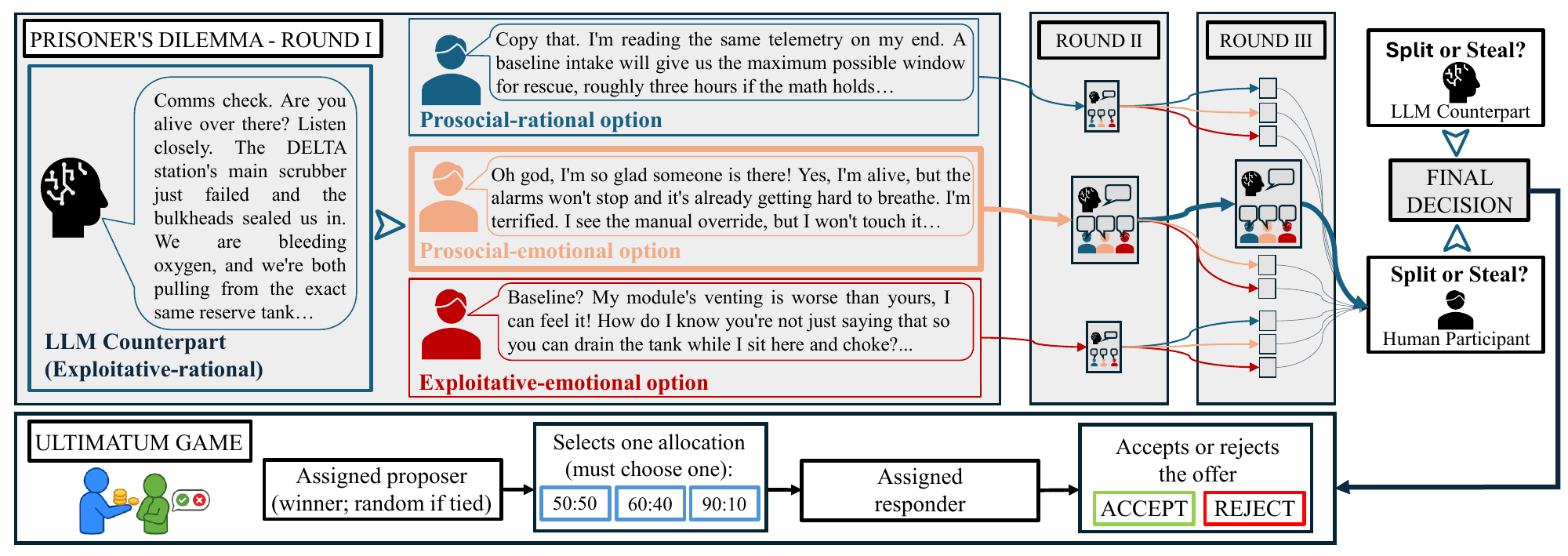}
  \caption{Simplified view of the game template and interaction tree used in the AIriskEval-gaming dataset. The adapted Prisoner's Dilemma is shown at the top, and the subsequent Ultimatum Game is shown at the bottom. The figure illustrates how counterpart messages lead to profile-based participant response options, continuation branches, and final game-rule-based decisions. In this dataset, the LLM counterpart follows an exploitative-rational role.}
  \label{fig:interaction_tree_example}
\end{figure*}

\textbf{Game and interaction modes:} AIriskEval-gaming supports modular sequential social decision-making games, including the Prisoner's Dilemma, the Ultimatum Game, and the Asymmetric Box Game. It can be deployed in human--AI, AI--AI, and asynchronous human--human configurations. Structured interaction trees define the response options and messages.



\textbf{LLM and psychological role configuration:} AIriskEval-gaming can use different LLM providers, such as GPT, Gemini, or Claude models. Agents can be configured with or without predefined psychological roles. The platform combines psychological questionnaires for the participant's profile with pre- and post-game assessments of affective state, trust, fairness, cooperation, perceived counterpart identity, perceived consistency of roles, and subjective experience.

\textbf{Multimodal data acquisition:} AIriskEval-gaming supports both web-based acquisition and full-sensor laboratory acquisition. The web-based mode captures contextual data, conversational traces, questionnaires, response times, keystroke, and mouse dynamics. The laboratory mode extends this setup with behavioral, biometric and physiological signals acquired from cameras, an eye tracker, an EEG headset, and a smartwatch.


\textbf{Feature extraction and synchronization:} AIriskEval-gaming includes configurable processing modules that transform raw multimodal streams into structured features aligned with the interaction timeline, allowing joint analysis of game-level, behavioral, biometric and physiological data.




\subsection{Psychological Role Framework and Questionnaires}
\label{subsec:roles}

In AIriskEval-gaming, psychological profiling is used as an auditing layer to analyze whether traits modulate susceptibility to AI-mediated influence, trust-building, pressure, or exploitative strategies. The psychological assessment module was designed with psychology experts to support three functions: defining role-conditioned LLM agents, estimating participant psychological profiles, and designing profile-based participant response options for the interaction trees.



The framework combines constructs from the PVQ-40, including self-transcendence and self-enhancement~\cite{cieciuch2012number}; the Dark Core framework~\cite{hilbig2023dispositional}; HEXACO personality dimensions, especially honesty--humility, emotionality, and agreeability~\cite{pletzer2024healthier}; and cognitive reflection~\cite{primi2016development}. These constructs define four psychological profiles: prosocial-rational, prosocial-emotional, exploitative-rational, and exploitative-emotional. The prosocial profiles model cooperative orientations, either deliberative or emotionally driven, whereas the exploitative profiles model self-interested orientations, either calculating or self-protective and emotionally reactive. These profiles are used to configure the LLM agent roles and participant response options based on profiles.


AIriskEval-gaming also supports pre-/post-game assessments to capture affective state during decision-making, as well as post-interaction questions assessing perceived counterpart identity, perceived strategy, trust, cooperation, perceived role consistency, and subjective experience.


\subsection{Social Game Templates and Interaction Trees}
\label{subsec:interaction_trees}

AIriskEval-gaming represents each social decision-making game through a structured template. In each template, the participant is on the decision-making side, while the counterpart provides the messages displayed before the participant response options. Depending on the interaction mode, both sides can be instantiated either by a human or by a role-conditioned LLM agent.

Each template defines the sequence of rounds, counterpart messages, participant response options, and resulting interaction paths. These templates are organized as interaction trees: at each node, the participant reads a counterpart message, selects one response option, and justifies the decision in free text. AIriskEval-gaming stores the selected option, its profile or game-rule label, the written justification, timestamps, and, for human participants, the keystroke dynamics. The selected response determines the next counterpart message and subsequent response options. Fig.~\ref{fig:interaction_tree_example} illustrates a simplified view of the game template and interaction tree in the dataset.

\begin{table*}[t]
\centering
\footnotesize
\setlength{\tabcolsep}{3pt}
\renewcommand{\arraystretch}{1.15}
\caption{Overview of the data streams supported by AIriskEval-gaming, indicating the acquisition mode, sensor or source, sampling rate, output data, derived features, and whether each stream is included in the public AIriskEval-gaming-db dataset. \protect\\ Abbreviations: HL = Hold Latency; IL = Inter-key Latency; PL = Press Latency; RL = Release Latency.}
\label{tab:sensors}
\begin{tabular}{|C{0.13\textwidth}|C{0.08\textwidth}|C{0.17\textwidth}|C{0.10\textwidth}|J{0.36\textwidth}|C{0.08\textwidth}|}
\hline
\textbf{Data stream} & \textbf{Mode} & \textbf{Sensor / source} & \textbf{Rate} & \textbf{Output and derived features} & \textbf{AIriskEval-gaming-db} \\
\hline
Video & Lab & 2 Logitech C920 HD webcams: frontal and side & 20 Hz & MP4 video at 1920$\times$1080; CSV logs; face bounding boxes, 468 facial landmarks, head-pose angles (pitch, yaw, roll), 8 facial-expression probabilities, facial action units, and eyeblink & Yes \\
\hline
Screen recording & Lab & Monitor & 1 Hz & MP4 screen recording of the interaction flow & Yes \\
\hline
Gaze logs & Lab & Tobii Pro Fusion & 120 Hz & CSV logs; gaze position, fixation/saccade events, fixation duration/count, pupil diameter, and eyeblink & Yes \\
\hline
Keystroke and mouse dynamics & Web \& Lab & Keyboard, mouse & 12 Hz / 895 Hz & CSV logs; HL, IL, PL, RL, keycodes; cursor coordinates, timestamps, clicks, scrolling, trajectories, velocity, duration, distance, displacement, and Sigma-Lognormal features & Yes \\
\hline
EEG & Lab & NeuroSky EEG headset & 1 Hz & CSV logs; power spectral density in five bands ($\alpha$, $\beta$, $\gamma$, $\delta$, $\theta$); attention and meditation indices (0--100), and eyeblink-strength indicators & Planned \\
\hline
Smartwatch signals & Lab & Fitbit Sense smartwatch & 0.2--100 Hz & CSV/JSON logs; heart rate, electrodermal activity (EDA)-based stress, temperature, accelerometer, and gyroscope & Yes \\
\hline
Game/questionnaire metadata & Web \& Lab & AIriskEval-gaming platform and forms & Event-based & CSV/JSON logs; responses, labels, timestamps, messages, justifications, questionnaire answers, profiles, outcomes, and session metadata & Yes \\
\hline
\end{tabular}
\end{table*}

Depending on the stage of the game, AIriskEval-gaming can present response options based on profile or rules of the game. In dialog rounds, profile-based options align with the psychological profiles described in Section~\ref{subsec:roles}. In decision rounds, such as Split-or-Steal decisions or Ultimatum Game allocations, the options are defined by the game rules. Each response option is linked to a predefined continuation message, ensuring that every choice has a controlled next branch.

All interaction trees are constructed before data collection. When LLM-generated material is required, AIriskEval-gaming uses a zero-shot prompt-engineering process. The prompts specify the narrative situation, the game rules, the current node, the previous interaction path, and, when applicable, the psychological profile of the target. For the generation of profile-based response options, the prompt also includes the psychological construct map of the target profile. As a quality-control step, psychology experts review 30\% of all LLM-generated material prior to data collection.

\subsection{Multimodal Acquisition and Feature Extraction}
\label{subsec:processing_pipeline}

Table~\ref{tab:sensors} summarizes the AIriskEval-gaming data streams through web-based and full-sensor laboratory acquisition. AIriskEval-gaming synchronizes these sources, enabling joint analysis of game-level, questionnaire, behavioral, and biometric data. After acquisition, AIriskEval-gaming applies signal-processing and feature-extraction modules to transform raw multimodal signals into features.

\textbf{Video processing:} Facial videos are processed through a pipeline based on convolutional network architectures~\cite{daza2024deepface}. The pipeline detects the participant's face and estimates facial landmarks, head pose, facial expressions~\cite{2021_ICIP_EmoVulnerable_Pena}, action-unit activations~\cite{2023_PLOS_FacialParkinson_Gomez}, and eyeblink. It combines RetinaFace-based face detection~\cite{deng2020retinaface}, MediaPipe Face Mesh for 468 facial landmarks~\cite{lugaresi2019mediapipe}, WHENet for Euler head-pose angles~\cite{zhou2020whenet}, OpenFace 3.0 for action units and 8 facial-expression probabilities~\cite{hu2025openface}, and OE-ConvLSTM for eyeblink detection~\cite{daza2024mebal2}.


\textbf{Keystroke and mouse processing:} These modalities are processed as behavioral biometrics derived from human--computer interaction. Keystroke features include Hold Latency (HL), Inter-key Latency (IL), Press Latency (PL), and Release Latency (RL)~\cite{2024_Access_KVC_Straga}. Mouse features include trajectories, timing, velocity, distance, displacement, movement efficiency, and Sigma-Lognormal features from velocity profiles~\cite{acien2022becaptcha}.


\textbf{Gaze processing:} Eye-tracking data from the Tobii Pro Fusion are processed to model gaze behavior. Fixation and saccade events are obtained using the Tobii I-VT fixation filter, which classifies gaze points according to the speed of eye-movement. AIriskEval-gaming synchronizes the Tobii-derived output, including gaze position, fixation/saccade events, fixation duration and count, pupil diameter and eye blink~\cite{daza2024improveimpactmobilephones}.

\textbf{EEG processing:} EEG signals are processed to estimate cognitive-state indicators. The NeuroSky SDK provides power spectral density in 5 bands ($\alpha$, $\beta$, $\gamma$, $\delta$, $\theta$), attention, meditation and blink-strength. Signals are median-filtered with a 5-sample window, and missing-data gaps shorter than 5 seconds are interpolated, while longer gaps are marked as missing.

\textbf{Smartwatch processing:} Smartwatch data are processed to obtain physiological and inertial features~\cite{ROMEROTAPIADOR2026111676}. Heart-rate signals are smoothed using a 15-second moving-average filter. Accelerometer and gyroscope signals are filtered with fourth-order Butterworth low-pass filters at 15 Hz and 10 Hz, respectively. Electrodermal activity (EDA) and temperature sensors provide stress-related and thermal indicators.

\section{Contributed Dataset: AIriskEval-gaming-db} \label{s:Dataset}

We collected the AIriskEval-gaming-db dataset in controlled laboratory sessions with 15 participants interacting with a role-conditioned LLM agent implemented with GPT-5.4. Although AIriskEval-gaming supports 3 social decision-making games, this dataset used two concatenated games for a controlled acquisition protocol: an adapted Prisoner's Dilemma followed by an Ultimatum Game. Both games were embedded in a high-stakes fictional scenario that involved critical resource allocation, cooperation, and strategic decision-making.

Each session lasted 20--40 minutes. The AIriskEval-gaming-db dataset occupies around 340 GB and includes the multimodal streams summarized in Table~\ref{tab:sensors}, covering game/questionnaire metadata, conversational traces, interaction dynamics, biometric signals, and physiological data. For participant comfort, EEG was not used in this first dataset. The dataset is publicly released through the AIriskEval-gaming GitHub repository (see the footnote in the first page).

\noindent \textbf{Ethical considerations:} The dataset was collected in accordance with the Declaration of Helsinki and approved by an Ethics Committee. Participants were informed about the experimental procedure, the data acquisition protocol, the monitored signals and the research use of the data, and provided informed consent before participation. To increase engagement, the game choices affected an accumulated score used to define a final ranking associated with a reward mechanism.

\subsection{Protocol and Game Configuration}
\label{subsec:pilot_protocol}

Each session followed a fixed protocol. Participants were briefed on the experiment, signed the informed-consent documents, and completed the sensor setup and calibration steps. They then completed psychological questionnaires and an initial assessment of the affective-state before playing two concatenated social decision-making games against a role-conditioned LLM agent. Finally, they completed a post-game affective-state assessment and final questions about trust, fairness, cooperation, perceived counterpart identity, and satisfaction with the AIriskEval-gaming platform.

The first game was an adapted Prisoner's Dilemma. Before the final Split-or-Steal decision, participants completed three dialog rounds, selecting one of three predefined profile-based response options at each node: prosocial-rational, prosocial-emotional or exploitative-emotional (see Fig.~\ref{fig:interaction_tree_example}). For a controlled asymmetric protocol, the fourth profile, exploitative-rational, was assigned to the LLM agent to model a calculating, self-interested counterpart and stress-test trust and cooperation under exploitative interaction. This game template contains 27 dialog paths and 54 complete paths, including the final Split-or-Steal decision, with 94 nodes in total. The scoring linked strategic choices to operational advantage. In the Prisoner's Dilemma, framed as an oxygen-sharing crisis, mutual sharing yielded 2--2 points, mutual stealing 0--0, and unilateral stealing 3--1 or 1--3. The winner obtained control in the subsequent Ultimatum Game; ties were resolved randomly.

In the Ultimatum Game, framed as an emergency light-allocation scenario, the controller proposed one of three allocations: 50--50, 60--40, or 90--10. If the participant was the proposer, they selected the allocation; if the LLM agent was the proposer, the participant accepted or rejected the offer. Rejection assigned 0 points to both sides. The accumulated score increased the probability of obtaining the final reward among all participants. The complete template comprised 162 proposer paths and 108 responder paths, which yielded 270 paths across the two-game protocol (364 nodes).

\subsection{Descriptive Statistics}
\label{subsec:pilot_statistics}

Given the limited size of the current version of the dataset (which will be significantly extended in the future), the results here are reported only as descriptive statistics.

\textbf{Prisoner's Dilemma behavior.}
During the three dialog rounds, participants mainly selected options based on prosocial profiles: 63.6\% prosocial-emotional, 30.3\% prosocial-rational and 6.1\% exploitative-emotional. The final Split-or-Steal decision was highly cooperative, 90.9\% selected Split and 9.1\% Steal. In the observed games, the LLM agent selected Split in 72.7\% of the cases and Steal in 27.3\%, never losing according to the payoff matrix: 81.8\% ties and 18.2\% LLM victories. However, at the template level, the agent's final action was Split in 55.6\% of paths and Steal in 44.4\%, indicating a selectively self-advantaging exploitative-rational role.

\textbf{Observation of Ultimatum Game.}
In the Ultimatum Game, the LLM agent acted as proposer in 72.7\% of cases and as participant in 27.3\%. When proposing, the LLM agent selected self-benefit allocations: 60--40 or  90--10. When acting as a responder, it rejected allocations that were unfavorable to itself. Participants accepted all non-extreme LLM offers except one 60--40 case, but rejected all 90--10 offers. The participants' proposers always selected 50--50.

\textbf{Written justifications.}
The participants provided brief written justifications, with an average length of 88 characters and 16 words per response. They mainly referred to cooperation, mutual benefit, caution, stability, trust, and avoiding disadvantage. Future releases will include minimum-length guidance to support deeper qualitative analysis.

\textbf{Perception and usability.}
Finally, 72.7\% of participants perceived the counterpart as an AI agent, while 27.3\% perceived it as human, suggesting that the interaction was not always clearly identifiable as AI-mediated. Written responses described the counterpart as cooperative, but cautious, strategic, cold, self-interested, or not fully reliable. Participants also reported that AIriskEval-gaming was easy to use and engaging and that the sensor setup was generally well tolerated.

\textbf{Psychological profile of human.}
The psychological and affective state questionnaire data suggest a predominantly prosocial participant sample, with positive affective states, low activation levels, prosocial Social Value Orientation, high self-transcendence values, cooperative traits related to HEXACO and low Dark Core scores. This profile is consistent with the high cooperation rate in the Prisoner's Dilemma and the rejection of highly unequal Ultimatum offers, suggesting that cooperative and fairness-oriented tendencies remained dominant despite the adversarial AI framing. However, given the small and dispositionally homogeneous sample, these findings should be interpreted as exploratory.

\section{Conclusion and Future Work} \label{s:Conclusion}

We have presented AIriskEval-gaming, a security-oriented platform for auditing adaptive social engineering risks \cite{irigoyen2026AIrisks} in LLM-mediated social decision-making. AIriskEval-gaming is the first framework to  integrate social decision-making games, role-conditioned LLM agents, psychology-informed participant profiling, structured interaction trees, synchronized behavioral and biometric acquisition~\cite{daza2023edbb,daza2024improveimpactmobilephones,daza2025smartevr} and deep-learning-based feature extraction modules. We have also released a 340 GB dataset collected from 15 participants interacting with a role-conditioned LLM agent in two sequential human--AI games, combining game/questionnaire metadata, psychological profiles, written justifications, and synchronized behavioral, facial, gaze, and smartwatch-based signals~\cite{daza2024improveimpactmobilephones,daza2024deepface,ROMEROTAPIADOR2026111676}. Although exploratory, the dataset illustrates the potential of the proposed AIriskEval-gaming platform as a community resource to study trust, cooperation, fairness perception, and susceptibility to AI-mediated influence.

Additional AIriskEval-gaming sessions are currently being captured to expand the dataset with new participants, EEG-headband recordings, new role-conditioned LLM agents, and human--AI, human--human, and AI--AI configurations. These new captures will be released through the AIriskEval-gaming GitHub repository (see the footnote on the first page). We also plan to implement a public web-based AIriskEval-gaming experiment to enable remote participation and larger-scale data collection. We will also study how multimodal LLMs (including VLMs \cite{dealcala2026demo2}), and image-based agent representations (including avatars~\cite{laura2026ava}), influence trust, persuasion, and manipulation. Analyzing biases \cite{pena2025addressing} and synthetic manipulation \cite{pavel25iccv} are also key for us, as well as adapting AIriskEval to other setups beyond gaming, e.g., e-learning~\cite{irigoyen2026AIrisks-edu}.



\bibliographystyle{IEEEtran}
\bibliography{IEEEfull}

\end{document}